\documentclass[11pt,a4paper]{article}

\usepackage[mathcal]{eucal}
\usepackage{color}
\usepackage{latexsym}
\usepackage{amsfonts}
\usepackage{amsmath}
\usepackage{amsthm}
\usepackage{amssymb}
\usepackage{mathrsfs}
\usepackage{graphicx}
\usepackage{algorithm}
\usepackage{algorithmic}

\newtheorem{theorem}{Theorem}[section]
\newtheorem{lemma}[theorem]{Lemma}

\newtheorem{observation}[theorem]{Observation}

\newcommand{\prbl}{\vspace*{0.2cm}}

\newenvironment{program}
	{\small  \begin{eqnarray*} } { \end{eqnarray*} }

\newlength{\programindent}
\newcommand{\prstyle}[1]{\mbox{\bf #1}}
\newcommand{\ind}[1]	{\hspace*{#1\programindent}}
\newcommand{\prline}[2]	{&& \ind{#1} #2 \\}

\newcommand{\Assign}[2]
	{#1 \leftarrow #2;}

\newcommand{\Whiledo}[1]
	{\prstyle{while~} #1 \prstyle{~do~}}

\newcommand{\Function}[2]
	{ \mbox{\rm #1}(#2)}
\newcommand{\Return}[1]
	{\prstyle{return~} #1;}

\begin{document}

\title{A constructive proof of the Lov\'asz Local Lemma}
\author{Robin A. Moser\thanks{Research is
   supported by the SNF Grant 200021-118001/1}
\\ \vspace{0.1cm} \\
Institute for Theoretical Computer Science\\
Department of Computer Science\\
ETH Z\"urich, 8092 Z\"urich, Switzerland\\
\texttt{robin.moser@inf.ethz.ch}
}
\date{October 2008}
\maketitle

\begin{abstract} 
The Lov\'asz Local Lemma \cite{EL75} is a powerful tool to prove the existence of combinatorial objects meeting
a prescribed collection of criteria. The technique can directly be applied to the satisfiability problem, 
yielding that a $k$-CNF formula in which each clause has common variables with at most $2^{k-2}$ other clauses is 
always satisfiable. All hitherto known proofs of the Local Lemma are non-constructive and do thus not provide a recipe
as to how a satisfying assignment to such a formula can be efficiently found. In his breakthrough paper \cite{Bec91}, Beck
demonstrated that if the neighbourhood of each clause be restricted to $\mathcal{O}(2^{k/48})$, a polynomial time algorithm for
the search problem exists. Alon simplified and randomized his procedure and improved the bound to $\mathcal{O}(2^{k/8})$ \cite{Alo91}. 
Srinivasan presented in \cite{Sri08} a variant that achieves a bound of essentially $\mathcal{O}(2^{k/4})$. In \cite{Mos08}, we 
improved this to $\mathcal{O}(2^{k/2})$. In the present paper, we give a randomized algorithm that finds a satisfying assignment to 
every $k$-CNF formula in which each clause has a neighbourhood of at most the asymptotic optimum of $2^{k-5}-1$ other clauses and
that runs in expected time polynomial in the
size of the formula, irrespective of $k$. If $k$ is considered a constant, we can also give a deterministic variant.
In contrast to all previous approaches, our analysis does not anymore invoke the standard non-constructive versions of the
Local Lemma and can therefore be considered an alternative, constructive proof of it.
\end{abstract}


\section{Introduction}

We use the notational framework introduced in \cite{Wel08}.
We assume an infinite supply of propositional \emph{variables}. A \emph{literal} $L$ is a variable $x$ 
or a complemented variable $\bar x$. A finite set $D$ of literals over pairwise distinct variables is 
called a \emph{clause}. We say that a variable $x$ \emph{occurs} in $D$ if $x \in D$ or $\bar x \in D$.
A finite set $F$ of clauses is called a \emph{formula} in CNF (Conjunctive Normal Form). We say that $F$
is a $k$-CNF formula if every clause has size exactly $k$. We write $\mbox{vbl}(F)$ to denote the set 
of all variables occurring in $F$. Likewise, $\mbox{vbl}(D)$ is the set of variables occurring in a clause
$D$. For a literal $L$, let $\mbox{vbl}(L) \in \mbox{vbl}(F)$ refer directly to the underlying variable.

A \emph{truth assignment} is a function $\alpha : \mbox{vbl}(F) \rightarrow \{0,1\}$ which assigns a boolean 
value to each variable. A literal $L = x$ (or $L = \bar x$ is \emph{satisfied} by $\alpha$ if $\alpha(x)=1$ 
(or $\alpha(x)=0$). A clause is \emph{satisfied} by $\alpha$ if it contains a satisfied literal and a formula 
is \emph{satisfied} if all of its clauses are. A formula is \emph{satisfiable} if there exists a satisfying
truth assignment to its variables.

Let $k \in \mathbb{N}$ and let $F$ be a $k$-CNF formula. The \emph{dependency graph} of $F$ is defined as
$G[F] = (V,E)$ with $V=F$ and 
$E = \{ \{C,D\} \subseteq V \; | \; C \ne D, \mbox{vbl}(C) \cap \mbox{vbl}(D) \ne \emptyset\}.$
The \emph{neighbourhood} of a given clause $C$ in $F$ is defined as the set 
$\Gamma_F(C) := \{ D \; | \; D \ne C, \mbox{vbl}(D) \cap \mbox{vbl}(C) \ne \emptyset \}$
of all clauses sharing common variables with $C$. It coincides with the set of vertices 
adjacent to $C$ in $G[F]$. 
The \emph{inclusive neighbourhood} of $C$ is defined to be $\Gamma_F^+(C) := \Gamma_F(C) \cup \{C\}$.

Suppose $U \subseteq \mbox{vbl}(F)$ is an arbitrary subset of the variables occurring in $F$, then we will denote
by $F_{(U)} := \{ C \in F \; | \; \mbox{vbl}(C) \cap U \ne \emptyset \}$ the subformula that is \emph{affected}
by these variables.

If $\alpha$ is any assignment for $F$, we write $\mbox{vlt}(F,\alpha)$ to
denote the set of clauses which are violated by $\alpha$.

The Lov\'asz Local Lemma was introduced in \cite{EL75} as a tool to prove the existence of
combinatorial objects meeting a prescribed collection of criteria. A simple, symmetric and uniform version of the Local Lemma can be
directly formulated in terms of satisfiability. It then reads as follows.
\begin{theorem}
{\bf \cite{EL75}}
If $F$ is a $k$-CNF formula such that all of its clauses $C \in F$ satisfy the property
$|\Gamma_F(C)| \le 2^{k-2}$, then $F$ is satisfiable.
\end{theorem}

The hitherto known proofs of this statement are non-constructive, meaning that they do not disclose
an efficient (polynomial-time) method to find a satisfying assignment. Whether there exists any
such method was a long-standing open problem until Beck presented in his breakthrough paper \cite{Bec91} 
an algorithm that finds a satisfying assignment in polynomial time, at least if $|\Gamma_F(C)| \le 2^{k/48}$. 
Using various guises of the
tools introduced by Beck, there have been several attempts to improve upon the exponent 
\cite{Alo91, Mos06, Sri08, Mos08},
but a significant gap always remained. In the present paper we will close that gap to the asymptotic optimum.
While the tools applied in the analysis are still derived from the original approach by Beck and also significantly
from the randomization and simplification contributed by Alon, the algorithm itself now looks substantially different.
An interesting new aspect of the present analysis is that the standard non-constructive proofs of the Local Lemma are
not invoked anymore. The algorithm we present and its proof of correctness can be therefore considered
a constructive proof of the Local Lemma, or at least of its incarnation for satisfiability. 
We conjecture that the methods
proposed can be seamlessly translated to most applications covered by the framework by Molloy and Reed in 
\cite{MR98}, however, this remains to be formally checked. Our main result will be the following.

\begin{theorem}
\label{mainthmrnd}
If $F$ is a $k$-CNF formula such that all clauses $C \in F$ satisfy the property
$|\Gamma_F^+(C)| \le 2^{k-5}$, then $F$ is satisfiable and there exists a randomized algorithm
that finds a satisfying assignment to $F$ in expected time polynomial in $|F|$ (independent
of $k$).
\end{theorem}

If we drop the requirement that the algorithm be of polynomial running time for asymptotically growing $k$, 
then we can also derandomize the procedure.

\begin{theorem}
\label{mainthmdet}
Let $k$ be a fixed constant. 
If $F$ is a $k$-CNF formula such that all 
clauses $C \in F$ satisfy the property
$|\Gamma_F^+(C)| \le 2^{k-5}$, then $F$ is 
satisfiable and there exists a deterministic 
algorithm that finds a satisfying assignment 
to $F$ in time polynomial in $|F|$.
\end{theorem}

In the sequel we shall prove the two claims.

\section{A randomized algorithm based on local corrections}

The algorithm is as simple and natural as it can get. Basically, we start with a random assignment, 
then we check whether any clauses are violated and if so, we pick one of them and sample another 
random assignment for the variables in that clause. We continue doing this until we either find a 
satisfying assignment or if the correction procedure takes too much time, we give up and restart 
with another random assignment. This very basic method turns out to be sufficiently strong for 
the case of formulas with small neighbourhoods, the only thing we need to do so as to make a running 
time analysis possible is to select the clauses to be corrected in a somewhat systematic fashion.

As in the theorem, let $F$ be a $k$-CNF formula over $n$ variables with $m$ clauses, such that
$\forall C \in F : |\Gamma_F^+(C)| \le d$, with $d := 2^{k-5}$. We impose an arbitrary, globally fixed ordering 
upon the clauses of $F$, let us call this the \emph{lexicographic ordering}. We define a recursive
procedure which takes the formula $F$, a starting assignment $\alpha$ and a clause $C \in F$ which
is violated by $\alpha$ as input, and outputs another assignment that arises from $\alpha$ by performing
a series of local corrections in the proximity of $C$.

%%%%%%%%%% ----- ALGORITHM LOCALLY CORRECT ------- %%%%%%%%%%%

\prbl
\begin{center}
\begin{minipage}{0.75\textwidth}
\hrule
\vspace*{-0.3cm}
\begin{program}
\prline{0}{\prstyle{function~} \Function{locally\_correct}{F, \alpha, C}}
\prline{1}{\Assign{\alpha}{\alpha \mbox{ with the assignments for vbl}(C) \; \mbox{replaced by random values (u.a.r.)}}}
\prline{1}{\Whiledo{\mbox{vlt}(\Gamma_F^+(C), \alpha) \ne \emptyset}}
\prline{2}{\Assign{D}{\mbox{lexicographically first clause in } \mbox{vlt}(\Gamma_F^+(C))}}
\prline{2}{\Assign{\alpha}{\Function{locally\_correct}{F, \alpha, D}}}
\prline{1}{\Return{\alpha}}
\end{program}
\vspace*{-1cm}
\hrule
%\vspace*{0.2cm}
\center{
\emph{Algorithm 2.1: recursive procedure for local corrections}
}
\end{minipage}
\end{center}
\prbl
\prbl
\prbl

%%%%%%%%%%%%%%%%%%%%%%%%%%%%%%%%%%%%%%%%%%%%%%%%%%%%%%%%%%%%%%%%

As you immediately notice, this recursion has the potential of running forever.
We will however see that long running times are unlikely to occur. If we are unlucky enough to encounter such a case,
we interrupt the algorithm prematurely. The following algorithm now uses the described recursive
subprocedure in order to find a satisfying assignment for the whole formula.

%%%%%%%%%%

\prbl
\prbl
\begin{center}
\begin{minipage}{0.75\textwidth}
\hrule
\vspace*{-0.2cm}
\begin{program}
\prline{0}{\prstyle{function~} \Function{solve\_lll}{F}}
\prline{1}{\Assign{\alpha}{\mbox{ an assignment picked uniformly at random from } \{0,1\}^{\mbox{vbl}(F)}}}
\prline{1}{\Whiledo{\mbox{vlt}(F,\alpha) \ne \emptyset}}
\prline{2}{\Assign{D}{\mbox{lexicographically first clause in } \mbox{vlt}(F,\alpha) }}
\prline{2}{\Assign{\alpha}{\Function{locally\_correct}{F,\alpha,D}}}
\prline{3}{\mbox{keep track of the number of recursive invocations done by locally\_correct;}}
\prline{3}{\mbox{if the number exceeds } \log m + 2, \mbox{ then abort the whole loop and}}
\prline{3}{\mbox{restart, sampling another $\alpha$.}}
\prline{1}{\Return{\alpha}}
\end{program}
\vspace*{-1cm}
\hrule
\center{
\emph{Algorithm 2.2: the complete solver}
}
\end{minipage}
\end{center}
\prbl
\prbl
\prbl

%%%%%%%%%%%

If the algorithm terminates, the result clearly constitutes a satisfying assignment. We however have
to check that the expected running time is polynomial. The remainder of the proof is to establish this
property.

In order to be able to talk about the behaviour
of the algorithm we need to control the randomness injected. Let us formalize the random bits used
in a way that will greatly simplify the analysis. Let us say that a total function 
$\mathcal{A} : \mbox{vbl}(F) \times \mathbb{N}_0 \rightarrow \{0,1\}$ is a \emph{table of assignments for $F$}.
Let us extend the notion to literals in the natural manner, i.e. $\mathcal{A}(\bar x, i) := 1-\mathcal{A}(x,i)$.
Furthermore call a total function $\alpha : \mbox{vbl}(F) \rightarrow \mathbb{N}_0$ an
\emph{indirect assignment}. Given a fixed table of assignments, an indirect assignment
automatically induces a standard truth assignment $^\star \alpha$ which is defined as
$^\star \alpha(x) := \mathcal{A}(x,\alpha(x))$ for all $x \in \mbox{vbl}(F)$. Let us now, just for
the analysis, imagine that the algorithm works with indirect assignments instead of standard
ones. That is, instead of sampling a starting assignment uniformly at random, $\Function{solve\_lll}{...}$
could sample a table of assignments $\mathcal{A}$ uniformly at random by randomly selecting each
of its entries. It will then hand over the pair $(\mathcal{A},\alpha)$ to $\Function{locally\_correct}{...}$,
where $\alpha$ is an indirect assignment this time which takes zeroes everywhere. Note that this
is equivalent since $^\star \alpha$ is uniformly distributed. Each time
the value of a variable is supposed to be resampled inside $\Function{locally\_correct}{...}$, we instead
increase its indirect value by one. Note that this equally completely corresponds to sampling a new
random value since the corresponding entry of the table has never been used before. In the sequel, we will
adopt this view of the algorithm and its acquisition of randomness.

We need to be able to record an accurate journal of what the algorithm does. A \emph{recursion tree} $\tau$ is
an (unordered) rooted tree together with a labelling $\sigma_\tau : V(\tau) \rightarrow F$ of each
vertex with a clause of the formula such that if for $u,v \in V(\tau)$, $u$ is the parent node of $v$, then
$\sigma_\tau(v) \in \Gamma_F^+(\sigma_\tau(u))$. We can record the actions of the recursive procedure
in terms of such a recursion tree, where the root is labelled with the clause the procedure was originally
called for and all descendant nodes representing the recursive invocations and carrying as labels the
clauses handed over in those. Let now a table of assignments $\mathcal{Á}$ be globally fixed and 
let $\alpha$ be any indirect assignment and $C \in F$ some clause violated under $^\star \alpha$. 
Suppose that $\Function{locally\_correct}{...}$ halts on inputs $\alpha$ and $C$. Then we say that the 
\emph{complete recursion tree} on the given input is the complete representation of the recursive process
up to the point where it returns. Even if the process does not return or does not return in the time we
allot, we can capture the tree representing the recursive invocations made in any intermediate step and
we call these \emph{intermediate recursion trees}.

Let $\tau$ be any recursion tree. The \emph{size} $|\tau|$ of $\tau$ is defined to be the number of vertices.
In order to simplify notation, we will write $[v] := \sigma_\tau(v)$
for any $v \in V(\tau)$ to denote the label of vertex $v$. Let us say that a variable $x \in \mbox{vbl}(F)$
\emph{occurs} in $\tau$ if there exists a vertex $v \in V(\tau)$ such that $x \in \mbox{vbl}([v])$. We
write $\mbox{vbl}(\tau)$ to denote the set of variables that occur in $\tau$.

We are now ready to make a first statement about the correctness of the algorithm.

\begin{lemma}
\label{lemreccorr}
Let $F$ and $\mathcal{A}$ be globally fixed. Let $\alpha$ be any indirect assignment and $C \in F$ a clause
violated under $^\star \alpha$. Suppose $\Function{locally\_correct}{...}$ halts on input $\alpha$ and $C$
and let $\tau$ be the complete recursion tree for this invocation. 
Then the assignment $\alpha'$ which the function outputs satisfies the subformula $F_{(\mbox{vbl}(\tau))}$.
\end{lemma}

\proof Assume that $\alpha'$ violates any clause $D \in F_{(\mbox{vbl}(\tau))}$. Suppose furthermore
that during the process we have recorded the ordering in which the recursive invocations were made.
Now let $v \in V(\tau)$ be the last vertex (according to that ordering) of which the corresponding 
label $[v]$ shares common variables with $D$. Since any invocation on input $[v]$ can only return once 
all clauses in $\Gamma_F^+([v])$ are satisfied, it cannot return before $D \in \Gamma_F^+([v])$ is. 
Since after that no changes of the assignments for the variables in $D$ have occured (by choice of $v$), 
$D$ is still satisfied when the function returns, a contradiction. \qed
\prbl

In the proof we had to assume having remembered
the ordering in which the process generated the vertices. However, from the statement of the lemma we
can now infer that this is not necessary since that ordering can be reconstructed by just looking at
the shape and the labels. For any recursion tree $\tau$, let us define the \emph{natural ordering} 
$\pi_\tau : V(\tau) \rightarrow [|V(\tau)|]$ to be the ordering we obtain by starting a depth-first
search at the root and at every node, selecting among the not-yet traversed children the one with the 
lexicographically first label (we use the notation $[n] := \{1,2,\mathellipsis,n\}$ for $n \in \mathbb{N}$). 
We claim the following.

\begin{lemma}
\label{lemrecord}
Let $\tau$ be any (intermediate) recursion tree produced by any (possibly non-terminating) call to 
$\Function{locally\_correct}{...}$. The ordering in which the process made recursive invocations
coincides with the natural ordering of $\tau$ and in particular, there is no node with two identically
labelled children.
\end{lemma}

\proof Note that a recursive procedure naturally generates a recursion tree in depth-first search
order. What we have to check is merely that the children of every node are generated according to
the lexicographic ordering of the their labels. We proceed by induction. The recursion tree 
$\tau$ has a root $r$ labelled $[r]$ and a series of children. Assume as induction hypothesis 
that the subtrees rooted at the children have the required property. We have to prove that on the 
highest invocation level where the loop checks for unsatisfied clauses in $\Gamma_F^+([r])$, no clause 
is either picked twice or is picked despite the fact that it precedes a clause already picked during
the same loop in the lexicographic ordering. Both possibilities are excluded by Lemma \ref{lemreccorr},
which readily implies that after a recursive invocation returns, the set of violated clauses is a strict
subset of the set of clauses violated before and in particular the clause we made the call for is satisfied
ever after. It can easily be verified that the fact that we allow intermediate trees does not harm.
\qed
\prbl

Lemma \ref{lemreccorr} also implies that the maximum number of times the outer loop 
(in $\Function{solve\_lll}{...}$) needs to be repeated is bounded by $\mathcal{O}(m)$ since
the total number of violated clauses cannot be any larger and each call to the correction
procedure eliminates at least one. Since we abort the recursion whenever it takes more
than logarithmically many invocations, clearly the whole loop terminates after a polynomial
number of steps. The only thing left to prove is that it is not necessary to abort and
jump back to the beginning more often than a polynomial number of times. In the sequel, we
will show that in the expected case, this has to be done at most twice.

\vspace*{-0.3cm}
\section{Consistency and composite witnesses}

Suppose that we fix an assignment table $\mathcal{A}$ and an indirect assignment
$\alpha$ and we call $\Function{locally\_correct}{...}$ for some violated clause $C$ and wait.
Suppose that the function does not return within the $\log(m)+2$ steps allotted and we
abort. Now let $\tau$ be the recursion tree that has been produced up to the time of
interruption. $\tau$ has at least $\log(m)+2$ vertices (up to some rounding issues 
it has exactly that many). Suppose that somebody were to be convinced
that it was really necessary for the process to take that long, then we can present them the
tree $\tau$ as a justification. By inspecting $\tau$ together with $\alpha$,
they can verify that for the given table $\mathcal{A}$, the correction could not have been
completed any faster. Such a certification concept will now allow us to
estimate the probability of abortion. We have to introduce some formal notions.

Let $v \in V(\tau)$ be any vertex and $x \in \mbox{vbl}([v])$ a variable that occurs there. 
We define the
\emph{occurrence index of $x$ in $v$} to be the number of times that $x$ has occurred before in
the tree, written
$$ \mbox{idx}_\tau(x,v) \; := \; | \{ v' \in V(\tau) \; | \;  
\pi_\tau(v') < \pi_\tau(v), \; x \in \mbox{vbl}([v'])  \} |. $$

If $\delta$ is any indirect assignment, we say that $\tau$ 
is \emph{consistent with $\mathcal{A}$ offset by $\delta$} if the property
$$ \forall v \in V(\tau) \; : \; \forall L \in [v] \; : \; \mathcal{A}(L, \mbox{idx}_\tau(\mbox{vbl}(L),v)+\delta(x) ) = 0$$
holds. Furthermore, let us define the \emph{offset assignment induced by $\tau$} as
$$ \delta_\tau(x) := | \{ v \in V(\tau) \; | \; x \in \mbox{vbl}([v]) \} |$$
for all $x \in \mbox{vbl}(F)$. Recall now what the recursion procedure does; it starts with a given indirect assignment
and performs recursive invocations in the natural ordering of the produced recursion tree and
in each invocation, the indirect assignments of the variables in the corresponding clause are
incremented. This immediately yields the following observation.

\begin{observation}
\label{obsrectrcons}
If $\tau$ is any (intermediate) recursion tree of an invocation of the recursive procedure for
a table $\mathcal{A}$ and an indirect starting assignment $\alpha$, then $\tau$ is consistent
with $\mathcal{A}$ offset by $\alpha$.
\end{observation}

Let a collection $W := \{\tau_1, \tau_2, \mathellipsis, \tau_t\}$ of recursion trees be given of which
the roots have pairwise distinct labels.
Consider an auxiliary graph $\mathcal{H}_W$ of which those trees are the vertices and two distinct vertices
$\tau_i, \tau_j$
are connected by an edge if the corresponding two trees share a common variable,
i.e. $\mbox{vbl}(\tau_i) \cap \mbox{vbl}(\tau_j) \ne \emptyset$. If $\mathcal{H}_W$ is connected,
then $W$ is said to be a \emph{composite witness} for $F$. The \emph{vertex set} of a composite witness
is defined to be $V(W) := \{(\tau,v) \; | \; \tau \in W, v \in V(\tau)\}$, i.e. it is the set of all vertices in
any of the separate trees, each annotated by its tree of origin. To shorten notation, we access the label
of such a vertex by writing $[w] := [v]$ for each $w=(\tau,v)\in V(W)$. The \emph{size} of a composite witness 
is defined to be $|V(W)|$.

There is a natural way of traversing the vertices in $V(W)$. Consider that each of the recursion
trees $\tau_i$ has a distinctly labelled root. Now order the trees according to the lexicographic ordering
of the root labels. Traverse the first tree according to its natural ordering, then the second one, and
so forth. We call this the \emph{natural ordering} for a composite witness and we write 
$\pi_W : V(W) \rightarrow [|V(W)|]$ to denote it. Let $v \in V(W)$ be any vertex of the composite
witness and $x \in \mbox{vbl}([v])$ any variable that occurs there. We define the
\emph{occurrence index of $x$ in $v$} to be the number of times that $x$ has occurred before in
the witness, written
$$ \mbox{idx}_W(x,v) \; := \; | \{ v' \in V(\tau) \; | \;  
\pi_W(v') < \pi_W(v), \; x \in \mbox{vbl}([v'])  \} |. $$
We say that a composite witness is \emph{consistent with $\mathcal{A}$} if the property
$$ \forall v \in V(W) \; : \; \forall L \in [v] \; : \; \mathcal{A}(L, \mbox{idx}_W(\mbox{vbl}(L), v)) = 0$$
holds. The definitions immediately imply the following.

\begin{observation}
\label{obsoffsetcons}
If $W = \{\tau_1, \tau_2, \mathellipsis, \tau_t\}$ is a composite witness with the recursion
trees ordered according to the lexicographical ordering of their root labels, then $W$ is consistent
with $\mathcal{A}$ if and only if for all $1 \le i \le t$, $\tau_j$ is consistent with $\mathcal{A}$ offset
by $\sum_{j<i} \delta_{\tau_j}$.
\end{observation}
\prbl

Moreover, note that for vertices $v \in V(W)$ and literals $L \in [v]$, the mapping
$$(v,L) \mapsto (\mbox{vbl}(L), \mbox{idx}_W(\mbox{vbl}(L),v))$$ is, by definition of $\mbox{idx}_W$,  an injection. This implies
that if we are given a fixed composite witness $W$ and we want to check whether $W$ is consistent with
a table $\mathcal{A}$, then for each pair $(v,L)$ to be checked, we have to look up one distinct entry
in the table. In total, we have to look up $k|V(W)|$ entries and $W$ can be consistent with $\mathcal{A}$
exclusively if each of those entries evaluates exactly as prescribed. This yields the following.

\begin{observation}
\label{obssmallprob}
If $\mathcal{A}$ is a random table of assignments where each entry has been selected uniformly at random from $\{0,1\}$, 
then the probability that a given fixed composite witness $W$ becomes consistent with $\mathcal{A}$ is exactly 
$2^{-k|V(W)|}$. 
\end{observation}
\prbl

In fact, we will now see that whenever the superintending procedure has to interrupt the recursion because it
has made too many invocations, then the bad luck at the origin of this behaviour can be certified by means of a 
large composite witness which occurs very rarely. We say that a composite witness is \emph{large} if it has 
size at least $\log m + 2$. We claim the following.

\begin{lemma}
Let an assignment table $\mathcal{A}$ be fixed and now run the loop in $\Function{solve\_lll}{...}$, i.e. starting
with the indirect all-zero assignment, repeatedly call the correction procedure on the lexicographically first clause 
violated by the current assignment and replace the assignment by the returned one. If the procedure fails, that is
if any of the local correction steps have to be interrupted because it needed at least $\log m + 2$ invocations, then
there exists a large composite witness for $F$ that is consistent with $\mathcal{A}$.
\end{lemma}

\proof Let $\tau_1, \tau_2, \mathellipsis, \tau_t$ be the sequence of recursion trees that certify the recursive
processes we started. Note that since the last step had to be interrupted, $\tau_t$ is intermediate and of size
at least $\log m + 2$. The other trees are complete recursion trees. Moreover, note that by Lemma \ref{lemreccorr},
every completed recursion process cannot have introduced any new violated clauses and therefore, the root labels
of the trees are ordered lexicographically.

Let now $\mathcal{H}$ be an auxiliary graph of which $\{\tau_1, \tau_2, \mathellipsis, \tau_t\}$ is the vertex
set and two trees are connected by an edge if there is a variable that occurs in both of them. Let us identify the
connected components of $\mathcal{H}$ and let $W \subseteq \{\tau_1, \tau_2, \mathellipsis, \tau_t\}$ be the connected
component of which $\tau_t$ is part. Note that all trees $\tau'$ that are not in $W$ do not have any variables in
common with any tree in $W$, therefore all their induced offset assignments $\delta_{\tau'}$ are zero on all 
of $\mbox{vbl}(W)$. By Observation \ref{obsrectrcons}, it follows that if $W = \{\tau_{i_1}, \tau_{i_2}, \mathellipsis, \tau_{i_r}=\tau_t\}$,
where the indices preserve the ordering, then for all $1 \le j \le r$, the tree $\tau_{i_j}$ is consistent
with $\mathcal{A}$ offset by $\sum_{j'<j} \delta_{\tau_{i_{j'}}}$. By Observation \ref{obsoffsetcons}, this yields the claim. \qed

Since we already know that a fixed composite witness is unlikely to become consistent with a random table, it just 
remains to count the number of composite witnesses that can possibly exist for $F$. This will yield a bound on the
probability that there is no large composite witness consistent with the table at all and thence also on the 
probability that each local correction will go through without the need for premature interruption.

\section{Encoding and counting composite witnesses}

In order to give an upper bound on the number of composite witnesses for a given formula, we will think about
how they can be encoded efficiently. The most obvious way of just encoding each recursion tree separately by
giving the shape of the tree and its labels does unfortunately not suffice. There are two properties of
witnesses that we can exploit. Firstly, the labellings of the separate trees are not arbitrary but follow the
rule that the label of a child is always either a neighbour in $G[F]$ to the label of the parent or identical
to that label. Moreover, the fact that the auxiliary graph that describes the interconnectivity of the recursion
trees inside the composite witness has to be connected yields that those trees have to lie in proximity of one
another such that only one root vertex for the whole composite witness will have to be stored.

More formally. Let $\mathcal{I}$ be the infinite rooted $(2d)$-ary tree. Note that every node in that tree
has at least twice as many children as there can be clauses in $\Gamma_F^+(C)$ for any $C \in F$. Let now
$R \in F$ be any clause. We will 'root', so to speak, $\mathcal{I}$ at $R$ and starting from there, 'embed' 
the nodes of $\mathcal{I}$ into $G[F]$ in the following fashion. Let $\sigma_R : \mathcal{I} \rightarrow G[F]$
be a labelling of the vertices of $\mathcal{I}$ such that the root $r$ is labelled $\sigma_R(r) = R$ and
whenever a node $v \in V(\mathcal{I})$ is labelled by $\sigma_R(v) = D \in F$, then the $2d$ children of $v$,
call them $c_1, c_2, \mathellipsis, c_{2d}$, are labelled as follows: for $1 \le i \le |\Gamma_F^+(D)| \le d$,
we label $\sigma(c_i) := (\mbox{the lexicographically } i-\mbox{th clause of } \Gamma_F^+(D))$.
For $d+1 \le i \le d+|\Gamma_F^+(D)| \le 2d$, we label $\sigma(c_i) := \sigma(c_{i-d})$ in the same
ordering. If there are remaining children then these are not labelled, that is the map $\sigma_R$ is partial. 
We will call the first $d$ children of every node the \emph{low children} and the remaining $d$ children the \emph{high children}
of that node.

Let us now say that a triple $\langle C, T, c \rangle$, where $C \in F$ is a clause, $T$ is a subtree of $\mathcal{I}$
containing the root and $c$ is a $2$-colouring (or $2$-partition) of the edges of $T$ is a \emph{witness encoding}. The
\emph{size} of a witness encoding is defined to be the number of vertices in the tree, $|V(T)|$.
We will see that we can reversibly encode each composite witness as such a triple and this will yield a bound that is strong
enough for our purposes.

\begin{lemma}
\label{lemencoding}
Let $u \in \mathbb{N}$.
There is an injection from the set of composite witnesses of size exactly $u$ for $F$ into the set of witness encodings of size $u$. 
The total number of distinct composite witnesses of size exactly $u$ that exist for $F$ is upper bounded
by $m \cdot 2^{u(k-1)}$.
\end{lemma}

\proof Let $W$ be any composite witness for $F$. To prove the claim, we have to transform $W$ into a triple $\langle C, T, c \rangle$
as described above in a reversible fashion. 

To begin with, we will show that any recursion tree $\tau$ can be encoded as a subtree of size $|\tau|$ of $\mathcal{I}$
which is rooted at any vertex $v$ of $\mathcal{I}$ which carries a label that also occurs in $\tau$, i.e. $\sigma_R(v) \in \mbox{Im}(\sigma_\tau)$.
The encoding is reversible for any choice of $R$ and any choice of such vertex $v$. We proceed as follows. Let $R$ and $v$
be chosen arbitrarily and let then $u \in V(\tau)$ be any vertex such that $\sigma_R(v) = [u]$. Let $u=u_0, u_1, \mathellipsis, u_j$ be the
succession of vertices that we encounter on the shortest path from $u$ to the root, where $u_j$ is the root of $\tau$. We now start
producing a subtree of $\mathcal{I}$ that is rooted at $v$. We add the unique high child $c_1$ of $v$ with $\sigma_R(c_1)=[u_1]$.
This child exists because obviously $[u_1]$ is a neighbour of $[u_0]$ in $G[F]$. To $c_1$ as the parent, we then add the unique
high child $c_2$ such that $\sigma_R(c_2)=[u_2]$, and so forth. We have built a path from $v$ to a vertex $c_j$ that
is labelled $[u_j]$, just like the root of $\tau$. Now, in the most natural fashion, attach to $v$ all subtrees of $u$ in such
a way that each descendant is labelled the same way in $\mathcal{I}$ as the corresponding vertex is labelled in $\tau$. Always
use the uniquely defined low children for this kind of embedding. Attach then to $c_1$ all the subtrees of $u_1$ except for
the one rooted at $u_0$ that we have already handled, using the same type of embedding. Do the same for $c_2, \mathellipsis, c_j$.
The result is a subtree of $\mathcal{I}$ of size $|\tau|$ which is rooted at $v$ with a one-to-one correspondence of its vertices
to the vertices of $\tau$ and such that every vertex is labelled under $\sigma_R$ in the same way as the corresponding vertex is
labelled in $\tau$. Note that this transformation is reversible. Whenever we are given such a subtree, we can start at its root
vertex and follow the unique high child of every node until not possible anymore. The vertex we end up at corresponds to the
original root of $\tau$. It is clear that from there, we can perform the same succession of 'rotations' backwards to completely 
reconstruct $\tau$.

Consider the composite witness now. Let $W = \{\tau_1, \tau_2, \mathellipsis, \tau_t\}$. 
Let $\mathcal{H}_W$ be, as usual, the auxiliary graph where those recursion
trees are the vertices and an edge exists between any two trees if a common variable occurs in both of them. By the definition of
composite witnesses, $\mathcal{H}_W$ is connected. This means that we can easily exhibit an ordering of the vertices of $\mathcal{H}_W$
such that each vertex in that ordering is connected to at least one of the vertices listed previously. W.l.o.g., let the ordering
$\tau_1, \tau_2, \mathellipsis, \tau_t$ have this property. It need of course not coincide with the lexicographic ordering of the
root labels.

Define both $C$ and $R$ to be the label of the root of $\tau_1$. This is going to be the root vertex of the whole construction. The labellings
of $\mathcal{I}$ will now be $\sigma_R$. We encode the recursion trees as subtrees of $\mathcal{I}$ in the ordering just devised
and we glue them together to form one large tree. Start with $\tau_1$ and encode it in the way described above as a subtree of
$\mathcal{I}$ which is rooted at the root of $\mathcal{I}$. This is possible because we have just labelled the root in this way.
Now assume we have added all trees up to $1 \le i \le t$, producing so far a subtree $T_i$ of $\mathcal{I}$ and now we have to add 
$\tau_{i+1}$. By the way we have ordered the trees,
there exists a variable $x \in \mbox{vbl}(\tau_{i+1}) \cap \mbox{vbl}(\tau_{j})$ for some $j < i+1$. Since the encoding as $T_i$ has
preserved all the labels, there is also at least one vertex in $T_i$ of which the label contains $x$. Let $g$ be a deepest such vertex
in $T_i$, i.e. $x \in \mbox{vbl}(\sigma_R(g))$ but no descendant of $g$ in $T_i$ has a label which contains $x$. Let $r_{i+1} \in V(\tau_{i+1})$ be any vertex in $\tau_{i+1}$
which contains $x$. Clearly, $[r_{i+1}]$ and $\sigma_R(g)$ are either identical or neighbours in $G[F]$, therefore there exists a unique 
low child $c_g$ of $g$ (in $\mathcal{I}$) which is labelled $\sigma_R(c_g) = [r_{i+1}]$. Note that by choice of $g$, $c_g$ is not present
in $T_i$. Use the procedure described before to produce
a representation of $\tau_{i+1}$ as a subtree of $\mathcal{I}$ which is rooted at $c_g$. Add this representation to $T_i$ and connect
the root $c_g$ to the parent $g$ by an edge. Mark this edge as a special \emph{glueing edge} by means of the colouring $c$. All other
edges are marked \emph{regular} by that colouring. Continue the process until all trees are added. Note that the tree $T := T_t$ produced
contains exactly $|V(W)|$ vertices.

It can easily be seen that the construction is reversible. If any such subtree, the root label $C=R$ and the colouring are given,
simply reconstruct all the labels of $\mathcal{I}$ as $\sigma_R$, then delete all glueing edges to produce a disjoint subforest
of $\mathcal{I}$. Each of the subtrees can be retransformed into the original recursion trees with the correct roots by the reversal
process described above. 

Now that we have found such an encoding, determining the numbers is not difficult anymore. There are $m$ choices to select
$C \in F$. According to a simple counting exercise by Donald Knuth in \cite{Knu69}, the number of rooted subtrees of size
exactly $u$ in $\mathcal{I}$ does not exceed $(2ed)^u$. There are less than $2^u$ distinct $2$-colourings of the edges of
such a tree. Therefore, the total number is upper bounded by $m(4ed)^u < m(16d)^u = m(2^{k-1})^u$, as claimed.
\qed

Since the composite witnesses are little in number, we can now infer that their occurrence is sufficiently unlikely.

\begin{lemma}
\label{nolargews}
Consider an assignment table $\mathcal{A}$ produced by selecting each entry from $\{0,1\}$
uniformly and independently at random. The probability that there exists in $F$ a large composite
witness which is consistent with $\mathcal{A}$ is at most 1/2.
\end{lemma}

\proof Let $X_u$ be the random variable that counts the number of composite witnesses of size
exactly $u$ which are consistent with $\mathcal{A}$. We have already derived that the probability that a fixed
composite witness of size $u$ becomes consistent with $\mathcal{A}$ is $2^{-ku}$. Combining this with
the previous lemma yields the following bound on the expected value of $X_u$:
$$ E(X_u) \le 2^{-ku} \cdot m \cdot 2^{u(k-1)} = m 2^{-u}. $$
If $X$ denotes the random variable that counts the total number of large composite witnesses which are 
consistent with $\mathcal{A}$, then for the expected value of $X$ we obtain
$$ E(X) = \sum_{u \ge \log(m)+2} X_u \le m \cdot 2^{-\log m} \sum_{u \ge 2} 2^{-u} = \frac{1}{2}. $$
If the number of such witnesses is at most 1/2 on average, then for at least half of the assignment tables,
there is no consistent such witness at all.  \qed

This concludes the proof of Theorem \ref{mainthmrnd}.

\section{A deterministic variant}

In this section we demonstrate that there is nothing inherently randomized to our procedure. If we assume $k$ to be
a constant, we can give a determinstic polynomial-time algorithm for the problem. The key idea will be that we enumerate
all possible large composite witnesses and then instead of sampling an assignment table at random and hoping that it will avoid
all of them, we deterministically search for a table that does. Since we can only enumerate a polynomial number of
composite witnesses, we have to more carefully check which of them are really relevant. The following lemma says that it
is not necessary to check arbitrarily large witnesses.

\begin{lemma}
\label{lemsubwitnesses}
Let $u \in \mathbb{N}$ and let $\mathcal{A}$ be a fixed table of assignments. If there exists a composite witness of size
at least $u$ for $F$ that is consistent with $\mathcal{A}$, then there exists also a composite witness of a size in the
range $[u,ku+1]$ which is equally consistent with $\mathcal{A}$.
\end{lemma}

\proof Assume that the claim is fallacious for a fixed value $u \in \mathbb{N}$. Then assume that $W$ is a composite witness
that constitutes a smallest counterexample, i.e. $W$ is a composite witness of size at least $u$ which is consistent with
$\mathcal{A}$, but there exists no witness of a size in the range $[u,ku+1]$ which is consistent with $\mathcal{A}$ and $W$ is smallest with this property. Clearly,
this implies that $W$ has size larger than $ku+1$.

Let $W = \{ \tau_1, \tau_2, \mathellipsis, \tau_t \}$ be the list of recursion trees contained in $W$, sorted according
to the lexicographic ordering of the root vertices. Now let us remove the very last vertex according to the natural ordering,
i.e. the vertex $\pi_W^{-1}(|V(W)|)$. If $\tau_t$ consists of a singleton vertex, this amounts to deleting $\tau_t$ from the
witness, in all other cases it amounts to removing the last vertex in the natural ordering from $\tau_t$. Let in both
cases $W^*$ be the modified collection of recursion trees. Now check the auxiliary graph $\mathcal{H}_{W^*}$ with $W^*$
being the vertex set and edges in case of common variables. Due to the deletion of the last vertex and therefore possibly of
a clause from the set of labels of $W^*$, the graph $\mathcal{H}_{W^*}$ might have fallen apart into several
connected components. However, considering that in the removed clause there were exactly $k$ variables and the variable sets
of the connected components are disjoint, the number of such components cannot exceed $k$. If we now select the largest
component $W'$ of $\mathcal{H}_{W^*}$, where large in this case means having the maximum number of vertices summed over the
trees in that component, then $W'$ is by definition again a composite witness and by choice of $W'$, $|V(W')| \ge (|V(W)|-1)/k \ge u$.
Note also that $W'$ is consistent with $\mathcal{A}$ because neither the removal of the last vertex in the natural ordering,
nor the removal of any trees covering variable sets disjoint from the set of variables covered by $W'$ can influence
the consistency property.

In total, we have found a composite witness $W'$ which is strictly smaller than $W$ and also consistent with $\mathcal{A}$.
Either the size of $W'$ is in the range $[u,ku+1]$, a contradiction, or it is a smaller counterexample, a contradiction as
well. \qed

The lemma implies that if we want to check whether a given assignment table $\mathcal{A}$ has the potential to provoque a premature
interruption of some correction step, then it suffices to check that no composite witness with a size in the range 
$[\log m + 2, k(\log m + 2) + 1]$ exists that is consistent with $\mathcal{A}$. The number of composite witnesses with a size in
that range is, by Lemma \ref{lemencoding}, bounded by
$$\sum_{j=\log m + 2}^{k(\log m + 2) + 1} m \cdot 2^{j(k-1)} < ((k-1)(\log m + 2) + 2) \cdot m \cdot 2^{(k(\log m + 2) + 1)(k-1)}$$
which is in turn bounded by a polynomial in $m$. This polynomially large set of critical composite witnesses can obviously
be enumerated by a polynomial time algorithm.

Assume now that we do not simply want to check tables for consistency with certain witnesses but we would like to directly
produce an assignment table with which no large composite witness is consistent. Note that during the algorithm, no variable's
indirect assignment is incremented more than $m \cdot (\log m + 2)$ times and therefore at most this number of rows in the
assignment table is used. Let us then define boolean variables $z_{i,x}$ for all $i \in [0, m \cdot (\log m + 2)]$ and all
$x \in \mbox{vbl}(F)$ to represent the entry $\mathcal{A}(x,i)$ of the table. If $W$ is any fixed composite witness, then
we can in the obvious way translate the consistency property for $W$ into a clause $C_W$ over the variables $z_{i,x}$ such
that a truth assignment $\beta$ to the variables $z_{i,x}$ violates $C_W$ if and only if the corresponding table with
$\mathcal{A}(x,i) = \beta(z_{i,x})$ for all $i$ and $x$ is consistent with $W$. Clearly, the clause $C_W$ has size exactly
$k|V(W)|$.

Let now a determinstic algorithm enumerate all composite witnesses for $F$ with a size in the range 
$[\log m + 2, k(\log m + 2) + 1]$, for each of them, generate a corresponding clause and collect all those clauses
in a CNF formula $G$ of length polynomial in $m$. By the same calculation as in the proof of Lemma \ref{nolargews}, 
the expected number of violated clauses in $G$ if we sample a truth assignment $\beta$ for $G$ uniformly at random, is smaller
than $1/2$. It is well-known that a formula with this property can be solved by a polynomial-time determinstic
algorithm using the method of conditional expectations (see, e.g., \cite{Bec91}). 
We can therefore obtain, deterministically and in polynomial time,
an assignment $\beta$ that satisfies $G$. The values of $\beta$ provide us with a corresponding assignment table
$\mathcal{A}$ with which no composite witness with a size in the range $[\log m + 2, k(\log m + 2) + 1]$ and by Lemma
\ref{lemsubwitnesses} therefore no large composite witness at all is consistent. If we now invoke our algorithm, replacing
all random values by the fixed table $\mathcal{A}$, we are guaranteed that no interruption will occur and
all corrections will go through to produce a satisfying assignment of $F$.

This concludes the proof of Theorem \ref{mainthmdet}.

\section*{Acknowledgements}

Many thanks go to Dominik Scheder for various very helpful comments 
and to my supervisor Prof. Dr. Emo Welzl for the continuous support.

\end{document}